\documentclass[aps,superscriptaddress,twocolumn,showpacs,prb,floatfix]{revtex4} %% groupedaddress,
\usepackage{graphicx,rotating,subfigure,amsmath,amsfonts,amssymb,delarray}

\newcommand{\e}{\text{e}}
\newcommand{\im}{\text{i}}
\def\l{\left}
\def\r{\right}
\def\12{\frac{1}{2}}

\begin{document}
% You should use BibTeX and apsrev.bst for references
\bibliographystyle{apsrev}

%\preprint{}

%% \title{Thermodynamics of the One-Dimensional Extended Hubbard Model at Half-Filling}
\title{The Half-Filled One-Dimensional Extended Hubbard Model: Phase diagram and Thermodynamics}

%% Magnetic susceptibility of Sr$_2$Cu$_{1-x}$Pd$_x$O$_3$ and other doped
%%  one-dimensional antiferromagnets}
%% Curie-like behavior of the susceptibility in doped one-dimensional
%%  antiferromagnets}
%% Magnetic susceptibility for doped one-dimensional antiferromagnets
% Optional argument for running titles on pages
%\title[]{}


% other informatio
\author{S. Glocke}
\affiliation{Bergische Universit\"at Wuppertal, Fachbereich Physik, 42097
  Wuppertal, Germany}
\author{A. Kl\"umper}
\affiliation{Bergische Universit\"at Wuppertal, Fachbereich Physik, 42097
  Wuppertal, Germany}
\author{J. Sirker}
\affiliation{Max-Planck-Institute for Solid State Research, Heisenbergstr.~1,
  70569 Stuttgart, Germany}

\date{\today}

\begin{abstract}
  We study the thermodynamics of the one-dimensional extended Hubbard model at
  half-filling using a density-matrix renormalization group method applied to
  transfer matrices. We show that the various phase transitions in this system
  can be detected by measuring usual thermodynamic quantities like the
  isothermal compressibility and the uniform magnetic susceptibility. For the
  isothermal compressibility we show that universal crossing points exist
  which allow to accurately determine the line where the charge gap vanishes.
  By studying in addition several correlation functions, we confirm the
  existence of a phase with long-range dimer order (bond order) which has been
  a matter of debate for several years.  According to our calculations this
  phase is located in a narrow region between the spin-density and
  charge-density wave phases up to a tricritical point which we estimate to
  be at $U_t=6.7\pm 0.2$, $V_t =3.5\pm 0.1$. Our results for the phase diagram
  are in good agreement with the most recent zero-temperature density-matrix
  renormalization group study, however, they disagree in some important
  aspects from the most recent Quantum-Monte-Carlo study.
  %% Arguments about stabilization of BOW in extended region in d>1 ??
\end{abstract}
%%\pacs{05.10.Cc}
%%\pacs{05.70.-a}
%\keywords{}

\maketitle

\section{Introduction}
Understanding the effects of competing interactions and the associated quantum
phase transitions in models of strongly correlated electron systems is still
one of the main issues of modern condensed matter physics. In one dimension
correlation effects are particularly strong and a number of analytical and
numerical tools have been developed for this case. This has led to intense
work on such systems in the last decades. One of the seminal models in this
context is the extended Hubbard model
\begin{widetext}
\begin{equation}
\label{Ham}
H = -t\sum_{j,\sigma}\l(c^\dagger_{j,\sigma} c_{j+1,\sigma} +h.c.\r)
+U\sum_j\l(n_{j,\uparrow}-\frac{1}{2}\r)\l(n_{j,\downarrow}-\frac{1}{2}\r) 
+ V\sum_j(n_j-1)(n_{j+1}-1)
-\frac{h}{2}\sum_j(n_{j,\uparrow}-n_{j,\downarrow})-\mu\sum_j n_j\; .
\end{equation} 
\end{widetext}
Here $c^\dagger_{j,\sigma}$ ($c_{j,\sigma}$) is a creation (annihilation)
operator for an electron with spin $\sigma=\uparrow,\,\downarrow$ at site $j$,
$n_{j,\sigma} = c^\dagger_{j,\sigma} c_{j,\sigma}$, and $n_j= n_{j,\uparrow} +
n_{j,\downarrow}$. $t$ is the amplitude of a nearest neighbor hopping, $U \geq
0$ an on-site and $V\geq 0$ a nearest-neighbor Coulomb repulsion. A possible
additional magnetic field is denoted by $h$ and the chemical potential by
$\mu$. Here we will concentrate on the half-filled case $\mu=0$. In the
following, we will only keep the hopping amplitude $t$ where it is necessary
for the sake of clarity and set $t=1$ otherwise.

In the strong coupling limit, $U,V\gg t$, it is easy to see by simple
energetical considerations that two different ground states exist: For $U<2V$
the system is an insulator with long-range charge density wave (CDW) order
whereas for $U>2V$ a state with quasi-long-range spin density wave (SDW) order
forms. The transition between these two phases in the strong coupling limit is
first order.\cite{Hirsch_EHM,Bari,vanDongen} In the weak coupling limit,
$U,V\ll t$, the model can be studied using bosonization and g-ology.
\cite{NakamuraPRB,Solyom} In this framework one finds again a phase transition
between the SDW and CDW phase at $U=2V$. In the spin sector this transition is
driven by an operator which turns from marginally irrelevant in the SDW phase
to marginally relevant in the CDW phase. The spin gap therefore opens up
exponentially slowly and the transition in the spin sector is of
Kosterlitz-Thouless (KT) type. In the charge sector, on the other hand, there
is a relevant operator in both phases leading to a charge gap. The amplitude
of this operator vanishes only at the transition line $U=2V$ so that the
charge gap disappears.\cite{NakamuraPRB} The transition in the charge sector
is therefore second order. Already from the strong and weak coupling
approaches it is clear that a point $(U_t,V_t)$ in the intermediate coupling
regime must exist where the order of the phase transition changes.

In the last few years the extended Hubbard model has attracted renewed
attention because it has been suggested that the phase diagram obtained by the
weak coupling g-ology approach and strong-coupling perturbation theory might
not be complete. Nakamura pointed out first that there is no symmetry
requiring the lines in $U$,$V$-parameter space, where the marginal operator in
the spin sector changes sign and where the relevant operator in the charge
sector vanishes, to coincide.\cite{NakamuraJPSJ,NakamuraPRB} The coupling
constants for these two operators do coincide in the standard g-ology approach
where they are calculated to first order in the interaction parameters.
However, they might differ once higher order corrections are taken into
account. This opens up the possibility for an intermediate phase. By
extracting the scaling dimensions related to the critical exponents of certain
correlation functions from finite size energy spectra, Nakamura indeed found a
phase with long-range dimer order in a small region between the SDW and CDW
phases. This phase is often called a bond-order wave (BOW) state. The
existence of such a phase around $U=2V$ in the weak coupling regime 
%% up to the tricritical point $(U_t,V_t)$ 
was supported by Quantum-Monte-Carlo (QMC) calculations
\cite{SenguptaSandvik,SandvikBalents} as well as by a g-ology approach where
the coupling constants have been calculated beyond leading
order.\cite{TsuchiizuFurusaki} However, in a first DMRG calculation
\cite{Jeckelmann} such a phase was only found {\it above} the tricritical
point $(U_t,V_t)$ and only directly at the first order transition line. A
later DMRG calculation,\cite{Zhang} on the other hand, qualitatively
confirms again the phase diagram as proposed by Nakamura.  Further evidence
for the existence of a BOW phase in the weak coupling regime was also provided
by a functional renormalization group analysis.\cite{TamTsai}

Although the most recent DMRG\cite{Zhang} and the most recent QMC
study\cite{SandvikBalents} agree that a BOW phase of finite extent does exist,
they disagree about the shape of this phase. Whereas in the phase diagram of
Ref.~\onlinecite{Zhang} the BOW phase ends at the tricritical point, it
extends beyond this point to larger values of $U,V$ in the phase diagram of
Ref.~\onlinecite{SandvikBalents}. The question whether the tricritical point
also marks the end of the BOW phase or is located on the BOW-CDW boundary
therefore remains an open issue.

In this work we will investigate the half-filled one-dimensional extended
Hubbard model using a density-matrix renormalization algorithm applied to
transfer matrices (TMRG). This numerical method allows it to calculate
thermodynamic properties of the model in the thermodynamic limit. We will
provide further evidence for the correctness of the phase diagram as first
proposed by Nakamura and give an estimate for the tricritical point
$(U_t,V_t)$. In particular, we will argue based on our numerical results that
the BOW phase ends at the tricritical point and does not extend to larger
values of $U,V$ in contrast to the findings in
Ref.~\onlinecite{SandvikBalents}. In the process, we will develop and discuss
criteria to identify the different phases and transition lines by considering
usual thermodynamic quantities like the uniform magnetic susceptibility, the
isothermal compressibility (charge susceptibility), and the specific heat.

Our paper is organized as follows: In Sec.~\ref{TMRG} we briefly introduce the
TMRG algorithm and compare results for the Hubbard model ($V=0$) with exact
results obtained by the Bethe ansatz. In Sec.~\ref{PD} we then present results
for a variety of thermodynamic quantities which allow us to determine the
phase diagram of the extended Hubbard model at half-filling. The last section
is devoted to our conclusions.
\section{The TMRG algorithm and the Hubbard model}
\label{TMRG}
The density-matrix renormalization group applied to transfer matrices (TMRG)
has been explained in detail in [\onlinecite{Peschel},
\onlinecite{GlockeKluemperSirker}, \onlinecite{SirkerKluemperEPL}]. Here we
only want to briefly discuss the most important aspects.  The TMRG algorithm
is based on a mapping of a one-dimensional quantum system to a two-dimensional
classical one by means of a Trotter-Suzuki decomposition.  In the classical
model one direction is spatial whereas the other corresponds to the inverse
temperature. For the classical system a so called quantum transfer matrix
(QTM) %- which is rather confusingly often called the quantum transfer matrix
%(QTM) - 
is defined which evolves along the spatial direction. At any non-zero
temperature the QTM has the crucial property that its largest eigenvalue
$\Lambda_0$ is separated from the other eigenvalues by a finite gap. The
partition function of the system in the thermodynamic limit is therefore
determined by $\Lambda_0$ only, allowing it to perform this limit exactly.
The Trotter-Suzuki decomposition is discrete so that the transfer matrix has a
finite number of sites or local Boltzmann weights $M$. The temperature is
given by $T\sim (\epsilon M)^{-1}$ where $\epsilon$ is the discretization
parameter used in the Trotter-Suzuki decomposition. The algorithm starts at
some high-temperature value where $M$ is so small that the QTM can be
diagonalized exactly. Using a standard infinite-size DMRG algorithm, sites are
then added to the QTM leading to a successive lowering of the temperature. A
source for a systematic error in these calculations is the finite
discretization parameter $\epsilon$. However, this only leads to errors of
order $\epsilon^2$ in all thermodynamic quantities considered in the
following. We will choose $\epsilon = 0.025$ or $0.05$ so that this systematic
error will only be of order $10^{-3}-10^{-4}$. Another error is introduced by
the truncation of the Hilbert space. This error will grow with decreasing
temperature and will finally make the calculations unreliable. Down to which
temperature the DMRG algorithm works will depend on the maximum number of
states $N$ kept in the truncated Hilbert space basis. The truncation error is
difficult to estimate.  We therefore start by comparing our TMRG results for
the Hubbard model ($V=0$) with exact results obtained by Bethe
ansatz.\cite{JuettnerKluemper_Hubbard} Within the TMRG algorithm nothing
changes fundamentally when we introduce the nearest-neighbor Coulomb repulsion
$V$ so that we expect a similar accuracy in this case.

\begin{figure}
\includegraphics*[width=0.9\columnwidth]{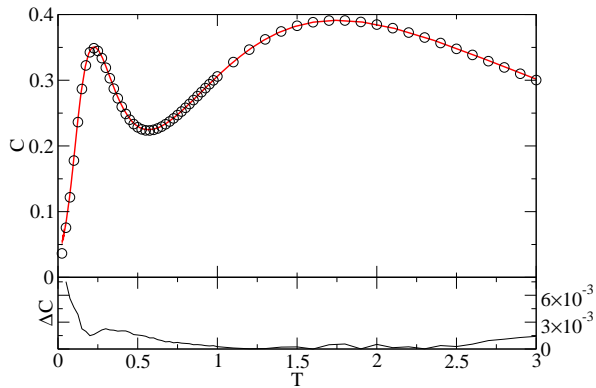}
\caption{(Color online) TMRG data for the specific heat $C$ of the Hubbard
  model at $U=8$ with $N=200$ states kept (red solid line) compared to Bethe
  ansatz data (circles) as a function of temperature $T$. The lower graph
  shows the error $\Delta C$ of the TMRG calculation.}
\label{Fig1}
\end{figure}

\begin{figure}
\includegraphics*[width=0.9\columnwidth]{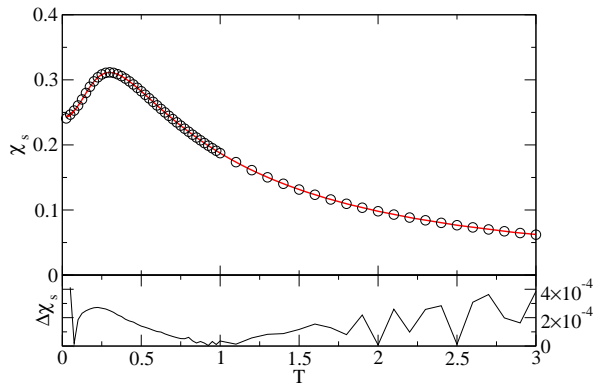}
\caption{(Color online) Same as Fig.~\ref{Fig1} for the magnetic
  susceptibility $\chi_s$.}
\label{Fig2}
\end{figure}

\begin{figure}
\includegraphics*[width=0.9\columnwidth]{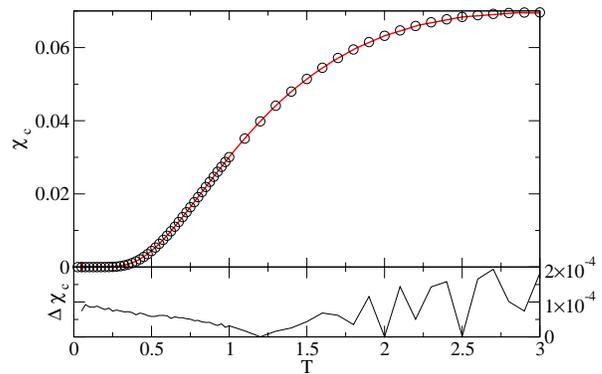}
\caption{(Color online) Same as Fig.~\ref{Fig1} for the charge susceptibility
  $\chi_c$.}
\label{Fig3}
\end{figure}

As an example, we consider the case $U=8$. Results with a similar accuracy are
also obtained for other $U$. Using the TMRG method, the free energy per site is
given by
\begin{equation}
\label{freeE}
f=-T\ln\Lambda_0 \;.
\end{equation}
The specific heat is then obtained by $C=-T\partial^2 f/\partial T^2$ and is
shown in  Fig.~\ref{Fig1}.
It is also easy to calculate the expectation values of local operators with
the TMRG algorithm. To obtain the magnetic susceptibility, $\chi_s$, the
expectation value $m\equiv \langle S^z\rangle = \langle n_\uparrow -
n_\downarrow\rangle/2$ is calculated in the presence of a small magnetic field
$\delta h\sim 10^{-2}$. The susceptibility is then given by $\chi_s=m/\delta
h$ and shown in comparison to the exact result in Fig.~\ref{Fig2}. 
Similarly, the isothermal compressibility (charge susceptibility), $\chi_c$, is
obtained by applying a small chemical potential $\delta\mu$ and is shown in
Fig.~\ref{Fig3}.
For the spin and charge susceptibilities $\chi_s,\,\chi_c$ the error does not
exceed $5\times 10^{-4}$ down to temperatures $T\approx 0.05$. For the specific
heat $C$, the errors are an order of magnitude larger because a second
numerical derivative has to be calculated.
\section{The phase diagram of the extended Hubbard model at half-filling}
\label{PD}
To investigate the phase diagram we will consider a number of different
thermodynamic quantities like magnetic susceptibilities, compressibilities,
specific heats, and expectation values of local operators. Furthermore, we
will study the behavior of correlation lengths which can be obtained within
the TMRG method by calculating next-leading eigenvalues of the
QTM.~\cite{Peschel,SirkerKluemperEPL} Depending on the required accuracy and
the temperature regime we want to access the basis for the truncated Hilbert
space will consist of $N=200-400$ states.

We start with the strong coupling limit where the two existing phases and the
first order phase transition between these phases are well understood. We then
derive an estimate for the tricritical point where the first order line ends.
Next we discuss the considerably more complicated weak coupling regime and
present the phase diagram as obtained by TMRG. Finally, we will address the
controversial question whether or not the BOW phase ends at the tricritical
point. Throughout, we will discuss in how far one can identify the different
phases and phase transitions by studying only easily measurable thermodynamic
quantities like the specific heat, magnetic susceptibility and
compressibility.
\subsection{Strong coupling}
\label{SC}
\begin{figure}
\includegraphics*[width=0.9\columnwidth]{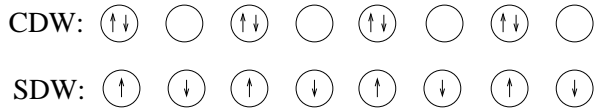}
\caption{The two ground states in the strong coupling limit $U,V\gg t$. The
  state in the first line is a CDW state where every second site is doubly
  occupied, whereas the state in the second line is a state with every site
  singly occupied. Virtual hopping processes induce a quasi long-range SDW
  order for this state.}
\label{Fig4}
\end{figure}

\begin{figure}
\includegraphics*[width=0.9\columnwidth]{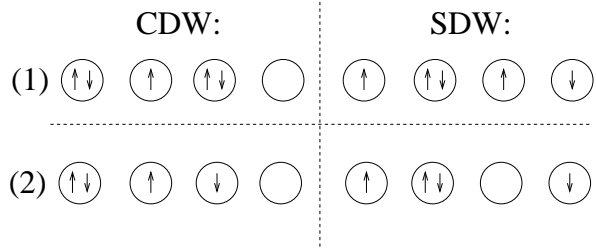}
\caption{Left column: (1) CDW state with one particle added, (2) CDW state
  with one double occupancy broken. Right column: (1) SDW state with one
  particle added, (2) SDW state with one double occupancy.}
\label{Fig_exc}
\end{figure}

%% Discuss Hubbard first, then finite V
In the strong coupling limit, $U,V\gg t$, the ground state energy can be
systematically expanded in terms of the hopping parameter $t$.  In lowest
order, the hopping can be completely neglected. Then, depending on the ratio
$U/V$, two different ground states are possible. These states are depicted in
Fig.~\ref{Fig4}.
The energy of the CDW state is then given by $E_{CDW}^0= LU/4-LV$ with $L$
being the number of lattice sites. The SDW state has energy $E_{SDW}^0=-LU/4$.
The two energies as a function of $U,V$ cross at $U=2V$ resulting in a first
order phase transition.  As usual, in second order in $t$, virtual hopping
processes lead to an effective antiferromagnetic coupling of Heisenberg type
for the spins in the SDW state with coupling constant $J=
2t^2/(U-V)$.\cite{Hirsch_EHM,vanDongen} This state therefore has a charge gap
but no spin gap and algebraically decaying spin correlation functions. The CDW
state, on the other hand, has a charge and a spin gap.  Excitations for the
CDW and SDW state, ignoring hopping processes, are shown in Fig.~\ref{Fig_exc}.
In lowest order perturbation theory, the energies of the excited states
depicted in Fig.~\ref{Fig_exc} are given by $E_{CDW}^1 = E_{CDW}^0 - U/2+2V,\,
E_{CDW}^2 = E_{CDW}^0-U+3V$ for the excited CDW states, and
$E_{SDW}^1=E_{SDW}^0+U/2,\, E_{SDW}^2=E_{SDW}^0+U-V$ for the excited SDW
states. Excitation (1) in the CDW state is a charge excitation, whereas the
breaking of a double occupancy - excitation (2) - is a spin excitation. If we
separate the two single spins in this excited state we obtain an excitation
energy $E_{CDW}^0 + 2(-U/2+2V)$, i.e., each single spin contributes $-U/2+2V$.
In thermodynamic data the activated behavior will be characterized by the
energy of a single excitation irrespective of whether these excitations appear
in pairs or not. In the strong coupling limit, it follows that at the
transition line charge and spin gap as obtained from thermodynamic data are
expected to be equal $\Delta_s=\Delta_c =U/2$ and that both gaps increase
linearly $\sim 2V$ away from the transition line. In the SDW phase, excitation
(1) is also a charge excitation and has a lower energy than excitation (2).
The charge gap in the SDW phase is therefore given by $\Delta_c=U/2$ and is
independent of $V$.

\begin{figure}
\includegraphics*[width=0.9\columnwidth]{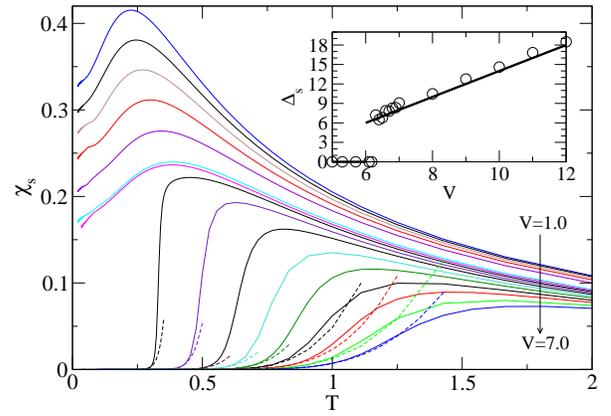}
\caption{(Color online) Magnetic susceptibilities (solid lines) for $U=12$ and
  $V=1.0,2.0,\cdots,6.0,6.1,6.2,\cdots,7.0$ as a function of temperature $T$.
  The dashed lines are fits according to Eq.~(\ref{compress2}). The inset
  shows the spin gap $\Delta_s$ extracted from those fits (circles) as a
  function of $V$. The solid lines in the inset denote the theoretical result
  in the strong coupling limit.}
\label{Fig5}
\end{figure}

 In Fig.~\ref{Fig5}, TMRG results for the spin susceptibility $\chi_s$ and the
spin gap $\Delta_s$ at $U=12$ are shown.
If a gap $\Delta$ exists, the dispersion of the elementary excitations is
given by $\epsilon(k)\sim \Delta + k^2/(2m)$ with some effective mass $m$. It
is then easy to see that the corresponding susceptibility will show activated
behavior
\begin{equation}
\label{compress2}
 \chi(T) \sim \frac{\exp(-\Delta/T)}{\sqrt{T}}
\end{equation}
at temperatures $T\ll \Delta$. Using this function to fit the numerical data,
we are able to extract the spin gap $\Delta_s$. As shown in the inset of
Fig.~\ref{Fig5} the behavior of $\Delta_s$ as a function of $V$ at
$U=12$ is already reasonably well described by the strong coupling limit,
i.e., there is no spin gap up to $V\approx U/2$, then $\Delta_s$ jumps to
approximately $U/2$ and then increases linearly with slope $2$.

\begin{figure}
\includegraphics*[width=0.9\columnwidth]{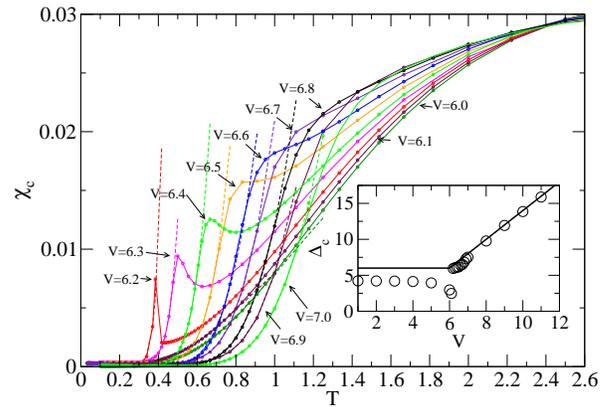}
\caption{(Color online) Charge susceptibilities (circles) for $U=12$ and
  $V=6.0,6.1,\ldots,7.0$ as a function of temperature $T$. The lines are a
  guide to the eye. The dashed lines are fits according to
  Eq.~(\ref{compress2}). The inset shows the charge gap $\Delta_c$ extracted
  from those fits (circles) as a function of $V$. The solid lines in the inset
  denote the theoretical result in the strong coupling limit.}
\label{Fig6}
\end{figure}

Similarly, we show TMRG results for the charge susceptibility $\chi_c$ and the
charge gap $\Delta_c$ at $U=12$ in Fig.~\ref{Fig6}.
The results obtained for the charge gap $\Delta_c$ are also already close to
the strong coupling limit, although the gap is a bit smaller than $U/2$ in the
SDW phase and it shows some $V$ dependence when the transition point is
approached.

\begin{figure}
\includegraphics*[width=0.9\columnwidth]{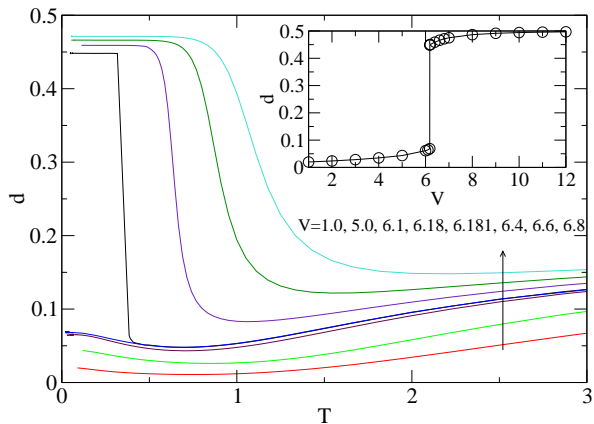}
\caption{(Color online) Double occupancy as a function of temperature for $U=12$ and
  different $V$. Inset: Extrapolated values of the double occupancy at zero
  temperature as a function of $V$.}
\label{Fig7}
\end{figure}

Another quantity which allows us to detect the phase transition and to determine
its order is the double occupancy 
\begin{equation}
\label{Def_double}
d=\langle
n_{j,\uparrow}n_{j,\downarrow}\rangle \; .
\end{equation}
In the strong coupling limit at zero temperature $d=0$ in the SDW state
and $d=1/2$ in the CDW state. In Fig.~\ref{Fig7} we show $d$ for $U=12$ and
various $V$.
In the extrapolated data for zero temperature some corrections to the strong
coupling limit are visible. $d$ is already nonzero in the SDW phase and
increases slightly with $V$. However, a jump in $d$ at $V\approx 6.18$ is
obvious. In the CDW phase $d$ continues to increase with $V$ and approaches
$1/2$ in the large $V$ limit.

\begin{figure}
\includegraphics*[width=0.9\columnwidth]{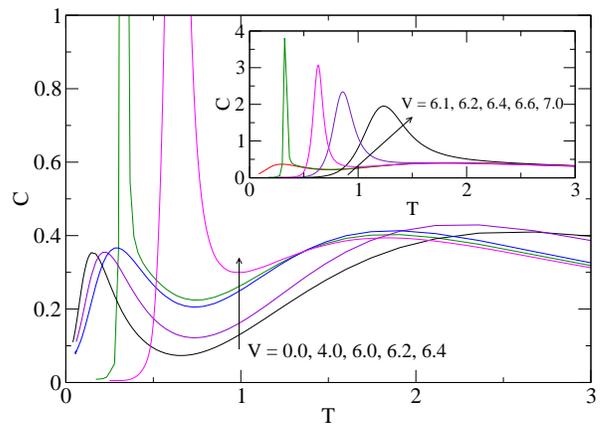}
\caption{(Color online) Specific heat as a function of temperature for $U=12$
  and different $V$. Inset: A sharp peak forms just above phase transition.}
\label{Fig15}
\end{figure}

%[We could/should add another Fig here showing the specific heat for $U=12$.
%Discussion of elementary excitations: two-peak structure for small $V$. I have
%only data for $U=8$.]
The specific heat shown in Fig.~\ref{Fig15} has two maxima for $U=12$ and
$V=0$. The lower and higher temperature maximum are due to spin and (gapped)
charge excitations, respectively.\cite{JuettnerKluemper} At low temperatures
only the gapless spin excitations do therefore contribute and conformal field
theory predicts 
\begin{equation}
\label{CFT_spec_Heat}
C=\frac{\pi}{3v_s}T \: .
\end{equation}
With increasing $V$ the spin velocity $v_s$ increases leading to a decreasing
slope and to a shift of the lower temperature maximum to higher temperatures.
At the same time the charge gap decreases leading to a shift of the higher
temperature maximum to lower temperatures. The behavior changes drastically
above the phase transition $V>V_c\approx 6.18$, because in the CDW phase the
spin excitations are now also gapped and the specific heat shows activated
behavior $C\sim e^{-\Delta/T}$ with $\Delta=\mbox{min}(\Delta_s,\Delta_c)$. The
emergence of a sharp peak for $V\gtrsim V_c$ can be understood as follows:
Because $\int_0^\infty C(T) dT = -e_0$ with $e_o$ being the ground state
energy, the area under the curve will be nearly unchanged when going from a
value just below the phase transition, say $V=6.1$, to a value just above the
transition, say $V=6.2$. In addition, also the high temperature behavior will
be almost unaffected by this small change in $V$. Hence the weight suppressed
by the gap at low temperatures will show up in a sharp peak just above the
gap. This is shown in the inset of Fig.~\ref{Fig15} and constitutes one
possibility to detect the first order transition easily from thermodynamic
data. 

\subsection{The tricritical point}
\label{MCP}
\begin{figure}[!ht]
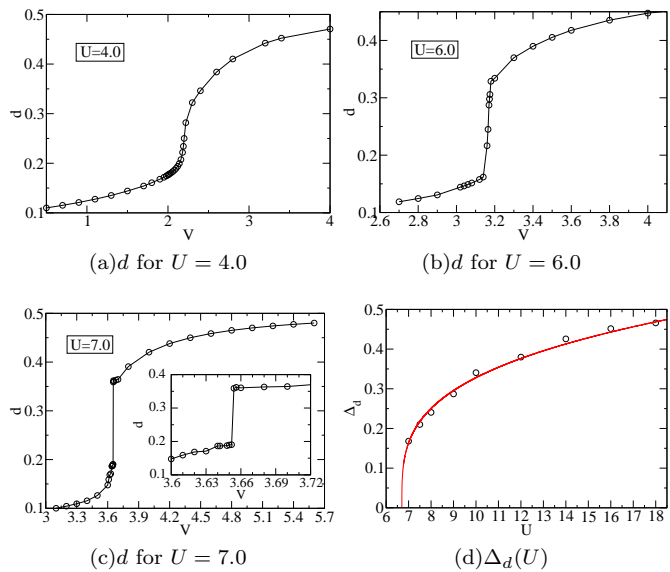

\centering
\subfigure[][$d$ for $U=4.0$]%in die letzte [] kommt text unters bild
{\label{fig:d_4U0} \resizebox{0.5\columnwidth}{!}
{\includegraphics*[]{d_4U0_0T.eps}}}%\quad
\subfigure[][$d$ for $U=6.0$]%$\alpha=-1,0;\ h=0;\; x=0,33348\pm2$]
{\label{fig9:d_6U0} \resizebox{0.5\columnwidth}{!}
{\includegraphics*[]{d_6U0_0T.eps}}}\\
\subfigure[][$d$ for $U=7.0$]
{\label{fig8:d_7U0} \resizebox{0.5\columnwidth}{!}
{\includegraphics*[]{d_7U0_0T_neu.eps}}}%\qquad
\subfigure[][$\Delta_d(U)$]
{\label{fig10:delta_d_U} \resizebox{0.5\columnwidth}{!}
{\includegraphics*[]{delta_d_U.eps}}}\\
\caption{(Color online) Extrapolated values of the double occupancy $d$ at
  zero temperature as a function of $V$ for different $U$. The inset of
  \subref{fig8:d_7U0} shows a zoom of the region where $d$ jumps.
  \subref{fig10:delta_d_U} Extrapolated TMRG data (circles)
  for the jump $\Delta_d$ in the double occupancy at the phase transition at
  zero temperature as a function of $U$. The line is a fit
  $\Delta_d=0.232(U-6.7)^{0.29}$.}
\label{fig:doppel}
\end{figure}

From the discussion in the introduction it is clear that the first order
transition line must end at some point $(U_t,V_t)$ because the phase
transitions at weaker couplings are expected to be continuous. We found that a
good criterion to determine this endpoint with the TMRG method is to study the
double occupancy $d$ (\ref{Def_double}). As shown for the case $U=12$ in
Fig.~\ref{Fig7}, $d$ as a function of $T$ shows a dramatically different
behavior depending on weather we choose a $V$ such that we are in the SDW
phase or a $V$ such that we are in the CDW phase. $d$ at a fixed $U$
extrapolated to zero temperature therefore shows a jump $\Delta_d$ as a
function of $V$ if the phase transition is first order. Reducing the on-site
repulsion $U$ we expect this jump to become smaller and smaller until it
disappears at $U_t$. For $U=7.0$ we can still detect a finite jump $\sim 0.17$
at $V\approx 3.65$ (see Fig.~\ref{fig8:d_7U0}) whereas $d$ as a function of
$V$ seems to be continuous for $U=6$ (see Fig.~\ref{fig9:d_6U0}).
%%%%%%%%%%%%%%%%%%%%
%%%%%%%%%%%%%%%%%%%%%
%\begin{figure}
%\includegraphics*[width=0.9\columnwidth]{d_7U0_0T_neu.eps}
%\caption{Extrapolated values of the double occupancy $d$ at zero
%  temperature as a function of $V$ for $U=7.0$. The inset shows a zoom of the
%  region where $d$ jumps.}
%\label{fig8:d_7U0}
%\end{figure}
%whereas $d$ as a function of $V$ seems to be continuous for $U=6$ (see
%Fig.~\ref{Fig9}). 
%\begin{figure}
%\includegraphics*[width=0.9\columnwidth]{d_6U0_4U0_0T.eps}
%\caption{Extrapolated values of the double occupancy $d$ at zero
%  temperature as a function of $V$ for $U=6$ and $U=4$ (inset).}
%\label{fig9:d_6U0}
%\end{figure}
%To determine the point $(U_t,V_t)$ more accurately we have plotted the jump
%$\Delta_d$ as a function of $U$ in Fig.~\ref{fig10:delta_d_U}. 
%\begin{figure}
%\includegraphics*[width=0.9\columnwidth]{delta_d_U.eps}
%\caption{(Color online) TMRG data (circles) for the jump $\Delta_d$ in the
%  double occupancy at the phase transition at zero temperature as a function
%  of $U$. The line is a fit $\Delta_d=0.232(U-6.7)^{0.29}$.}
%\label{fig10:delta_d_U}
%\end{figure}
To determine the point $(U_t,V_t)$ more accurately we have plotted the jump
$\Delta_d$ as a function of $U$ in Fig.~\ref{fig10:delta_d_U}. 
We can fit these data very well by a power law which leads us to the estimate
$U_t =6.7\pm 0.2$. Because the value for $U=7$ is least reliable, we also did
fits where this point was excluded. Similarly, we tried fits where the data
points for large $U$ were excluded. The results of the various fits lead to
the error estimate above. For each possible value of $U_t$ we can find $V_t$
with high accuracy. For the values of $U_t$ estimated above, we have $V_t =
3.5\pm 0.1$. Here the uncertainty in $V_t$ is not an error estimate but rather
means that $V_t\approx 3.4$ for $U_t=6.5$ and $V_t\approx 3.6$ for $U_t=6.9$.

\subsection{Weak coupling}
\label{WC}
%% Discuss Hubbard first, then finite V ?
The phase diagram in the weak coupling limit is more complicated than in the
strong coupling limit. Instead of a first order, we expect different
continuous phase transitions here. Theoretically, the weak-coupling limit can
be investigated by bosonization with the coupling constants of the operators
in the effective Hamiltonian being determined in first order in the
interaction parameters.  This method is also often termed
g-ology.\cite{Solyom} As usual in one-dimension, the charge and the spin
sector completely separate in the low-energy effective bosonic theory due to
the linearization of the excitation spectrum. In the charge sector at
half-filling, Umklapp scattering leads to a relevant interaction term in the
bosonic Hamiltonian which creates a charge gap. In the spin sector, on the
other hand, the leading interaction term corresponding to backward scattering
is only marginal. The amplitudes of both terms in the weak-coupling limit are
proportional to $U-2V$.\cite{NakamuraPRB} The system therefore has always a
charge gap except at $U=2V$ where the amplitude of the Umklapp scattering term
vanishes.  The charge gap at fixed $U$ near the phase transition behaves as
\begin{equation}
\label{charge-gap}
\Delta_c\sim |V-V_c|^\alpha
\end{equation}
with $\alpha>0$ being an interaction dependent critical exponent and
$V_c\approx U/2$ at weak coupling. This means that the transition in the
charge sector is second order.  In the spin sector at weak coupling the
backward scattering term is marginally irrelevant if $U>2V$ so that the spin
excitations are gapless. For $U<2V$ this term becomes marginally relevant and
a spin gap $\Delta_s$ appears. However, this gap only opens up exponentially
slow, i.e., for a fixed $U$ and $V\gtrsim V_c$ we expect
\begin{equation}
\label{spin-gap}
\Delta_s \sim \sqrt{V-V_c}\exp\{-\mbox{const}/(V-V_c)\}
\end{equation}
with $V_c\approx U/2$ at weak coupling.\cite{NakamuraKitazawa} The phase
transition in the spin sector is therefore of Kosterlitz-Thouless (KT) type.  

As Nakamura [\onlinecite{NakamuraJPSJ}] first noted, there is no symmetry
which fixes the amplitude of the Umklapp and backward scattering terms to be
the same. So although these amplitudes are identical to first order in the
interaction parameters, one would expect in general that they start to differ
once higher order corrections are taken into account. In this case an
additional phase between the SDW and CDW phases would occur. As already
outlined in the introduction different methods have given strong evidence that
an additional phase with BOW order does indeed exist although some controversy
about the extent of this additional phase
remains.\cite{NakamuraJPSJ,NakamuraPRB,Jeckelmann,SenguptaSandvik,Zhang,TsuchiizuFurusaki,TamTsai,SandvikBalents}

In the following, we will first develop a criterion which allows us to
determine the second order line where the charge gap closes with high
precision from thermodynamic data. Next, we will consider the KT-type
transition where the spin gap opens. Finally, we will provide some direct
evidence that the new phase has long-range BOW order at zero temperature and
does not extend beyond the tricritical point.

To determine the line in the $U,V$-phase diagram where the charge gap closes,
we consider the charge susceptibility $\chi_c$. 
% \begin{equation}
% \label{compress}
% \chi_c = \frac{\partial^2 f}{\partial \mu^2} = \frac{\partial \langle
%   n\rangle}{\partial \mu} \; . 
% \end{equation}
If a charge gap $\Delta_c$ exists, $\chi_c$ at temperatures $T\ll \Delta_c$
is described by Eq.~(\ref{compress2}). In the low temperature regime, $\chi_c$
therefore will be larger the smaller the charge gap is. According to
Eq.~(\ref{charge-gap}) we therefore expect the following behavior of
$\chi_c(T_0,V)$ at fixed $U$ and fixed low temperature $T_0$: If $V<V_c$ then
$\chi_c(T_0,V)$ increases with increasing $V$ whereas $\chi_c(T_0,V)$
decreases with increasing $V$ if $V>V_c$. 

\begin{figure}
\includegraphics*[width=0.9\columnwidth]{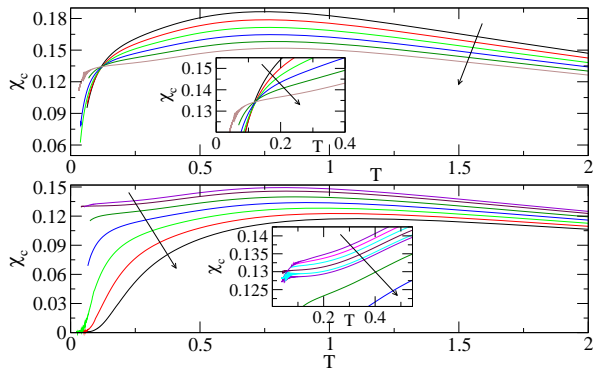}
\caption{(Color online) Charge susceptibility $\chi_c$ for $U=2$. Upper panel:
  $\chi_c$ for $V=0.5,0.6,\cdots,1.0$ (in arrow direction). The inset shows a
  zoom of the region around the crossing point at $T^*\approx 0.12$. Lower
  panel: $\chi_c$ for $V= 1.04,1.1,1.2,\cdots,1.6$ (main, in arrow direction)
  and $V=1.04,1.06,1.08,1.1,1.12,1.13,1.2,1.3$ (inset, in arrow direction).}
\label{Fig11}
\end{figure}

\begin{figure}
\includegraphics*[width=0.9\columnwidth]{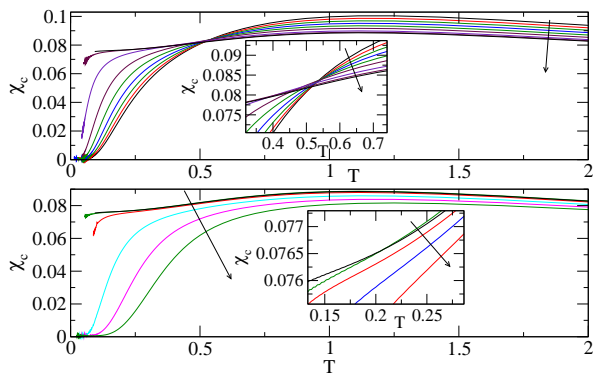}
\caption{(Color online) Charge susceptibility $\chi_c$ for $U=4$. Upper panel:
  $\chi_c$ for $V=1.5,1.6,\cdots,2.1,2.15,2.17$ (in arrow direction). The
  inset shows a zoom of the region around the crossing point at $T^*\approx
  0.54$. Lower panel: $\chi_c$ for $V= 2.16,2.17,2.2,2.3,2.4,2.5$ (main, in
  arrow direction) and $V=2.16,2.17,2.18,2.19,2.2$ (inset, in arrow
  direction).}
\label{Fig12}
\end{figure}

For high temperatures $T_0\gg 1$, on the other hand, $\chi_c(T_0,V)$ will
always decrease with increasing $V$ as can be easily seen from a high
temperature expansion. Up to second order in $1/T$ we find
\begin{equation}
\label{compress3}
 \chi_c(T\gg 1) = \frac{1}{2T}\l[1-\frac{1}{2T}(U/2+V)\r] \; . 
\end{equation} 
For $V<V_c$ we therefore have the situation that $\partial\chi_c/\partial V >
0$ for $T\ll 1$ and $\partial\chi_c/\partial V < 0$ for $T\gg 1$. The
compressibility curves for different $V<V_c$ at fixed $U$ therefore have to
cross at least at one point. For $V>V_c$, on the other hand, we have
$\partial\chi_c/\partial V < 0$ for high as well as for low temperatures so
that no crossing is expected.  The different behavior of the compressibility
curves for $V>V_c$ and $V<V_c$ is a very efficient criterion to determine $V_c$
as is shown in Figs.~\ref{Fig11}, \ref{Fig12} for the cases $U=2$ and $U=4$,
respectively.
From Fig.~\ref{Fig11}, lower panel, we see that the first curve crossing the
ones for larger $V$ is the one for $V=1.04$. We therefore find $V_c=1.05\pm
0.01$ for $U=2$. In principle, the critical point can be determined with this
method even more accurately. Similarly for $U=4$, the inset of the lower panel
of Fig.~\ref{Fig12} shows that the first curve crossing is $V=2.16$ which
leads to the estimate $V_c=2.165\pm 0.005$ in this case. Both critical values
are in good agreement with the most recent zero temperature DMRG
calculation.\cite{Zhang} Another interesting point is that in both cases the
curves for $V<V_c$ do not only cross but do so at one well defined point.
I.e., there is a well defined temperature $T^*$ where $\partial\chi_c/\partial
V \approx 0$ for all $V$. Similar well defined crossing points have also been
observed in other systems and other thermodynamic quantities, as for example,
the specific heat.\cite{Vollhardt, KemperSchadschneider}

\begin{figure}
\includegraphics*[width=0.9\columnwidth]{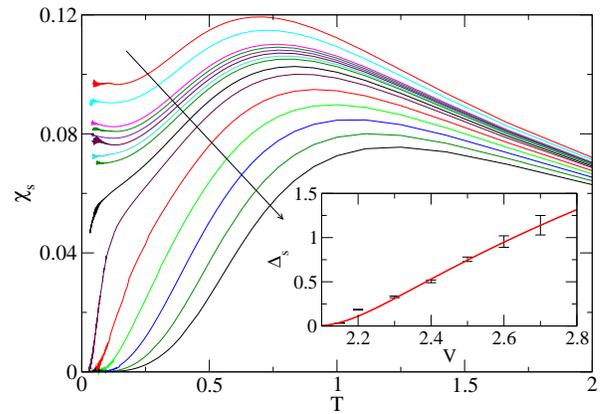}
\caption{(Color online) Spin susceptibility at $U=4$ for different
  $V=1.8,1.9,2.0,2.02,\cdots,2.1,2.15,2.2,2.3,\cdots,2.7$. Inset: Spin gap
  $\Delta_s$ at $U=4$ as a function of $V$. The gap is well fitted by
  $\Delta_s\sim 2.52\sqrt{V-2.02}\cdot \exp[-0.41/(V-2.02)]$.}
\label{Fig13}
\end{figure}

For the spin susceptibility, $\chi_s$, there is only a spin gap above the KT
transition. For $V>V^{KT}_c$ the temperature dependence of the spin
susceptibility is then again given by Eq.~(\ref{compress2}). $\partial
\chi_s/\partial V < 0$ for all temperatures so that the curves do not cross.
The same is true for $V<V^{KT}_c$: In the low-temperature limit the spin
sector is then described by conformal field theory and
\begin{equation}
\label{CFT1}
\chi_s(T=0) = \frac{1}{2\pi v_s} \; .
\end{equation}
The spin velocity $v_s$ increases with increasing interaction strength so that
again $\partial \chi_s/\partial V < 0$ for all temperatures. Therefore no
qualitative change happens at the transition line. In principle, one can try
to use the fact that there is universal scaling of certain ratios of
thermodynamic quantities in the conformal regime. The entropy is given by
Eq.~(\ref{CFT_spec_Heat}) so that
\begin{equation}
\label{CFT2}
\lim_{T\to 0} \frac{S}{T\chi_s} \equiv \frac{2\pi^2}{3} 
\end{equation}
is universal in the regime with gapless spin excitations.  However, these
formulas are only valid at temperatures $T\ll \Delta_c$. Because the spin gap
opens close to the point where the charge gap vanishes, this criterion turns
out to be useless for our numerical calculations. We therefore have to
determine the KT line by directly extracting the gap from the susceptibility
curves. As an example, we consider again the case $U=4$ (see
Fig.~\ref{Fig13}). 
For small $V$ the behavior is qualitatively consistent with Eq.~(\ref{CFT1})
whereas for large $V$ a spin gap is clearly visible. Fitting the low
temperature part of the curves where a gap is present using
Eq.~(\ref{compress2}) we can extract $\Delta_s$ as a function of $V$ as shown
in the inset of Fig.~\ref{Fig13}.
%\begin{figure}
%\includegraphics*[width=0.9\columnwidth]{Spin_gap_U4_0.eps}
%\caption{Spin gap $\Delta_s$ at $U=4$ as a function of $V$. The gap is well
%  fitted by $\Delta_s\sim 2.52\sqrt{V-2.02}\cdot \exp[-0.41/(V-2.02)]$.}
%\label{Fig13b}
%\end{figure}
Here, the error bars are obtained by varying the fit region. Another fit
according to Eq.~(\ref{spin-gap}) then yields $V_c^{KT} \approx 2.02\pm 0.06$
where the error estimate stems again from a variation of the fit region.
Within the estimated errors we therefore obtain strong evidence that $V_c\neq
V_c^{KT}$ for $U=4$, i.e., that we have two separate phase transitions.
Following this procedure to determine the second order and the KT transition
lines for other values of $U$, we obtain the phase diagram discussed in the
next subsection.
\subsection{Phase diagram}
\label{PD2}
\begin{figure}
\includegraphics*[width=0.9\columnwidth]{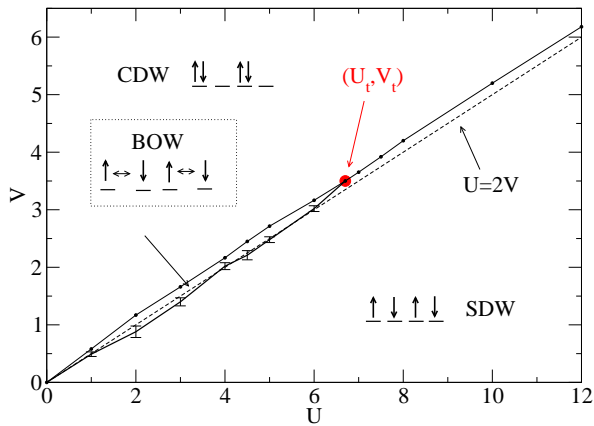}
\caption{(Color online) Phase diagram as obtained by TMRG. The dashed line
  denotes $U=2V$. The upper line describes the phase boundary of the CDW
  phase. The related error is always smaller than the symbol size. The error
  of the KT phase transition (lower line) is shown.}
\label{Fig14}
\end{figure}

%Discuss and show transition lines. 
Our phase diagram, shown in Fig.~\ref{Fig14}, is very similar to the one
obtained in the most recent zero-temperature DMRG study.\cite{Zhang} There is
a first order transition line for $(U,V)$ values above the tricritical point
$(U_t,V_t)$, separating the SDW and CDW phases. Below the tricritical point we
have a KT-type transition line where the spin gap opens and a second order
phase transition line where the charge gap disappears. The nature of the so
called BOW phase enclosed by the two transition lines is discussed in more
detail in the next subsection. There is some quantitative difference between
our study and Zhang's DMRG study\cite{Zhang} in the location of the
tricritical point though. We find $U_t=6.7\pm 0.2$, $V_t =3.5\pm 0.1$, whereas
he found $U_t\approx7.2$, $V_t \approx3.746$. Both values are considerably
larger than the ones found in QMC calculations, $U_t=4.7\pm
0.1$, $V_t =2.51\pm 0.04$ in Ref.~\onlinecite{SenguptaSandvik}, and
$U_t\in[5,5.5]$ in Ref.~\onlinecite{SandvikBalents}. We also note that our phase
diagram disagrees with that obtained in an earlier DMRG
calculation\cite{Jeckelmann} where the BOW phase was restricted to the first
order phase transition line (SDW-CDW) extending from {\it above} the
tricritical point estimated to be at $U_t=3.7\pm 0.2$ up to $U\approx 8$.

\subsection{Long-range BOW order and extent of the BOW phase}
From the phase diagram, Fig.~\ref{Fig14}, we see that the spin gap opens
starting from the SDW phase and only after that the transition into the CDW
phase occurs. From field theoretical considerations it is then expected that
the phase enclosed by these two transition lines is a Mott state with some
dimerization, also called a bond-order wave (BOW) state. Because such a
dimerization does not break any continuous symmetry, true long-range order at
zero temperature will occur even in one dimension. This means
that for the correlation function
\begin{equation}
\label{corrfunc}
F(r) = (-1)^r \l(\langle A_0 A_r\rangle - \langle A_r\rangle^2\r)
\end{equation}
with $A_r = S^z_r S^z_{r+1}$ or $A_r = \sum_\sigma
(c^\dagger_{r,\sigma} c_{r+1,\sigma} + h.c.)$ we have $\lim_{r\to\infty} F(r)
=\mbox{const} \neq 0$. With the TMRG algorithm there are different
possibilities to detect this order. First, next-leading eigenvalues of the QTM
allow it to calculate correlation lengths easily. In an asymptotic expansion
of a two-point correlation function with operator $O_r$ we obtain
\begin{equation}
\label{corrlengths}
\langle O_1O_r\rangle -\langle O_1\rangle\langle O_r\rangle = \sum_\alpha M_\alpha \e^{-r/\xi_\alpha} \e^{\im k_\alpha r}
\end{equation}
with correlation lengths $\xi_\alpha$ and wave vectors $k_\alpha$ given by
\begin{equation}
\label{corrlengths2}
\xi_\alpha^{-1} = \ln\l|\frac{\Lambda_0}{\Lambda_\alpha}\r| \quad , \quad
k_\alpha =\arg\l(\frac{\Lambda_0}{\Lambda_\alpha}\r) \; ,
\end{equation}
where $\Lambda_0$ is the largest eigenvalue of the QTM and $\Lambda_\alpha$
another eigenvalue. A correlation length obtained according to
Eq.~(\ref{corrlengths2}) will show up in the asymptotic expansion
(\ref{corrlengths}) if the corresponding matrix-element $M_\alpha$, which can
also be calculated with the TMRG
algorithm,\cite{SirkerKluemperEPL,SirkerKluemperPRB} is nonzero.  In the long
distance limit, the behavior of the correlation function will be determined by
the largest correlation length $\xi$ with nonzero matrix-element.

If the correlation function decays algebraically, this correlation length will
diverge like $\xi\sim 1/T$. If, on the other hand, the correlation function
decays exponentially even at zero temperature then $\xi$ stays
finite. Finally, for a correlation function showing true long-range order at
zero temperature the correlation length will diverge like 
\begin{equation}
\label{long-range}
\xi\sim \frac{\exp(\Delta/T)}{\sqrt{T}}
\end{equation}
where $\Delta$ is the gap for the corresponding excitations. 

\begin{figure}[!ht]
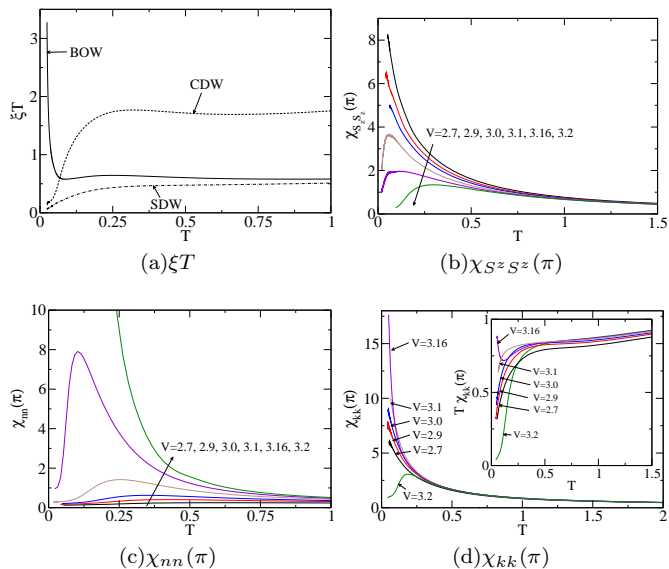

\centering
\subfigure[][$\xi T$]%in die letzte [] kommt text unters bild
{\label{Fig_corr} \resizebox{0.5\columnwidth}{!}
{\includegraphics*[]{corr_6U0_3V16.eps}}}%\quad
\subfigure[][$\chi_{S^zS^z}(\pi)$]
{\label{SDW} \resizebox{0.5\columnwidth}{!}
{\includegraphics*[]{sus_alter_Sz_6U0.eps}}}\\
\subfigure[][$\chi_{nn}(\pi)$]
{\label{CDW} \resizebox{0.5\columnwidth}{!}
{\includegraphics*[]{sus_alter_CDW_6U0.eps}}}%\qquad
\subfigure[][$\chi_{kk}(\pi)$]
{\label{BOW} \resizebox{0.5\columnwidth}{!}
{\includegraphics*[]{sus_alter_kin_6U0.eps}}}\\
\caption{(Color online) \subref{Fig_corr}: Leading SDW, CDW and BOW
  correlation lengths plotted as $\xi T$ for $U=6$ and $V=3.16$.
  \subref{SDW}-\subref{BOW}: alternating static susceptibilities for the
  longitudinal spin, the density, and the kinetic energy $k=\sum_\sigma
  (c^\dagger_{r,\sigma} c_{r+1,\sigma} + h.c.)$ for $U=6$ and different $V$,
  respectively.}
\label{fig:korrel_und_sus_alter}
\end{figure}

In Fig.~\ref{Fig_corr} we show, as an example, the leading SDW, CDW and BOW
correlation lengths for $U=6$ and $V=3.16$.
%%%%%%%%%%%%%%%%%%%%%%%%%%%%
%%%%%%%%%%%%%%%%%%%%%%%%%%%%
%\begin{figure}
%\includegraphics*[width=0.8\columnwidth,angle=270]{corr_6U0_3V16.ps}
%\caption{Leading SDW, CDW and BOW correlation lengths plotted as $T\xi$ for $U=6$ and $V=3.16$.}
%\label{Fig_corr}
%\end{figure}
Here the leading SDW and CDW correlation lengths stay finite whereas the BOW
correlation length diverges faster than $1/T$ indicating long-range BOW order
at zero temperature. 

Another possibility to detect the BOW order with the TMRG algorithm is to
calculate static susceptibilities
\begin{equation}
\label{static-susci}
\chi_{OO}(q)=\sum_r \e^{iqr}\int_0^\beta d\tau \langle O_0(0) O_r(\tau)\rangle
\end{equation}
again for some operator $O_r$. For true long-range order the corresponding
$\chi(q)$ will diverge exponentially with temperature, whereas $\chi(q)$ will
go to a constant (zero if the operator is conserved) for short-range order.
The situation is, however, complicated if the correlation function shows quasi
long-range order, i.e., decays algebraically. Here we want to consider the
case that only one sort of excitations is gapless, say the spin excitations.
From conformal field theory it is known that the corresponding algebraically
decaying correlation function in the long-distance limit $r\gg 1$ will behave
as
\begin{eqnarray}
\label{corr_CFT}
\langle O_0(0)O_r(\tau)\rangle &\sim& \l(\frac{2\pi
  T}{v}\r)^{2x}\exp\l[\frac{-2\pi Tx}{v}r\r] \\[0.2cm]
&\times & \exp[-\im kr]\exp[2\pi T\im(d^+ - d^-)\tau] \; . \nonumber
\end{eqnarray}
Here $v$ is the velocity of the elementary excitations, $x=d^+ + d^-$ the
scaling dimension, $d^\pm$ the conformal weights, and $k$ the characteristic
wave vector. The $\tau$-integral for the static susceptibility $\chi_{OO}(k)$ can
then be calculated explicitly and is given by 
\begin{equation}
\label{integral}
\int_0^\beta d\tau \exp[2\pi T\im(d^+ - d^-)\tau]
=\frac{\e^{2\pi\im(d^+-d^-)}-1}{2\pi\im T(d^+ - d^-)} \; .
\end{equation}
If the conformal spin $s=d^+ - d^-$ is a non-zero integer -- this is the case
for any type of particle-hole excitation -- the integral is zero and this part
of the correlation function does not contribute. If, on the other hand, $s=0$
then there is no time dependence in (\ref{corr_CFT}) and the integral
(\ref{integral}) yields just $1/T$. The static susceptibility in the case of
zero conformal spin will therefore scale as $\chi_{OO}(k)\sim T^{2x-2}$. In
particular, for the alternating part of the longitudinal spin-spin correlation
function we have $d^+ = d^- =1/4$ leading to $\chi_{S^zS^z}(\pi)\sim 1/T$.
Note, however, that for $x>1$ the long-distance asymptotics is no longer
sufficient to discuss the behavior for $T\to 0$. In this case $\chi_{OO}(k)\to\mbox{const}$
for a non-conserved operator in general as in the case of exponentially decaying
correlation functions discussed above.

\begin{figure}
\includegraphics*[width=0.9\columnwidth]{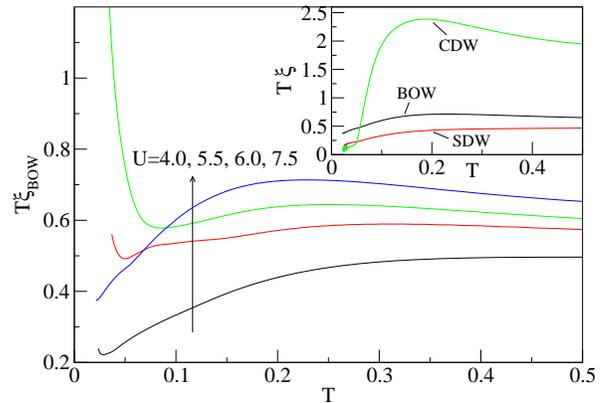}
\caption{(Color online) Leading BOW correlation lengths for $(U,V) =
  (4.0,2.14),(5.5,2.9),(6.0,3.19),(7.5,3.9)$. The inset shows the leading BOW,
SDW, and CDW correlation lengths for $(U,V)=(7.5,3.9)$.}
\label{Fig_add}
\end{figure}

In Figs.~\ref{SDW}, \ref{CDW}, \ref{BOW} we show for $U=6$ and different $V$
the alternating static susceptibilities for the longitudinal spin, the
density, and the kinetic energy, respectively.
%\begin{figure}
%\includegraphics*[width=0.9\columnwidth]{sus_alter_Sz_6U0.eps}
%\caption{$\chi_{S^zS^z}(\pi)$ for $U=6$ and different $V$.}
%\label{SDW}
%\end{figure} 
%\begin{figure}
%\includegraphics*[width=0.9\columnwidth]{sus_alter_CDW_6U0.eps}
%\caption{$\chi_{nn}(\pi)$ for $U=6$ and different $V$.}
%\label{CDW}
%\end{figure} 
%\begin{figure}
%\includegraphics*[width=0.9\columnwidth]{sus_alter_kin_6U0.eps}
%\caption{$\chi_{kk}(\pi)$ for $U=6$ and different $V$.}
%\label{BOW}
%\end{figure}
From Fig.~\ref{SDW} we conclude that a spin gap develops for $V\gtrsim
3.1$. However, for $V=3.1$ and $V=3.16$ there is still no long-range charge
order (see Fig.~\ref{CDW}), i.e., an intermediate phase does exist. In
Fig.~\ref{BOW} we see that at least for $V=3.16$ this phase has long-range BOW
order which is consistent with the correlation lengths shown in
Fig.~\ref{Fig_corr}. Fitting the BOW correlation length using
Eq.~(\ref{long-range}) we extract a rather small dimer gap $\Delta\sim
0.08$. For fixed $U$ the dimer gap is expected to decrease with decreasing $V$
so that possible long-range bond order is detected most easily close to the
transition into the CDW phase. In Fig.~\ref{Fig_add} the leading BOW
correlation lengths for several $U,V$-values just below this transition line are
shown. 
For $(U,V) = (4.0,2.14),(5.5,2.9)$, and $(6.0,3.19)$ the correlation lengths
diverge exponentially and we obtain the dimer gaps $\Delta \approx 0.01,
0.03$, and $0.08$, respectively. As expected, $\Delta$ decreases with
decreasing $U$ making it difficult to show the exponential divergence of the
BOW correlation length for $U<4$ because temperatures below $T\sim 10^{-2}$
are not easily accessible by the TMRG method. Nevertheless, it is clear that
the whole phase enclosed by the two transition lines shown in Fig.~\ref{Fig14}
must have long-range BOW order. For $(U,V)=(7.5,3.9)$, on the other hand, we
would expect $\Delta\gtrsim 0.1$ if BOW order does exist as found in
Ref.~\onlinecite{SandvikBalents} so that an exponential divergence should
already become obvious at $T\sim 0.1$. However, down to $T\approx 0.02$ we see
no indication of such a behavior, instead the BOW correlation length seems to
diverge exactly as $1/T$ indicating that we are in the SDW phase. This is
supported by the data in the inset of Fig.~\ref{Fig_add} showing that the
leading SDW correlation length also diverges like $1/T$ whereas the CDW
correlation length stays finite for $T\to 0$. Interestingly, the BOW
correlation length is larger than the SDW correlation length. We also
confirmed that for $(U,V)=(7.5,3.92)$ we are already in the CDW phase. We
therefore conclude that for $U=7.5$ no BOW phase exists. Instead, a direct first
order transition from the SDW to the CDW phase occurs.

\section{Summary and Conclusions}
\label{Con}
We studied the thermodynamics of the half-filled one-dimensional extended
Hubbard model using a TMRG algorithm. The focus was put on identifying the
various phase transitions by considering thermodynamic quantities which are
usually easy to measure like the uniform magnetic susceptibility, $\chi_s$, or
the isothermal compressibility, $\chi_c$. For strong coupling we calculated
the charge gap in the SDW as well as charge and spin gap in the CDW phase in
lowest order perturbation theory. The theoretical results were confirmed by
TMRG calculations of $\chi_s$ and $\chi_c$. In the weak coupling regime where
the phase transitions are continuous we showed that $\chi_c$-curves for a
fixed $U$ and different $V$ as a function of temperature cross in one well
defined point if measured in the SDW or BOW phase. In the CDW phase, on the
other hand, no crossing occurs. We used this criterion to determine the
boundary of the CDW phase with high accuracy. The KT transition line, on the
other hand, where the spin gap starts to open exponentially slowly is very
difficult to determine from thermodynamic data.  Universal scaling relations
obtained from conformal field theory for the magnetic susceptibility and the
specific heat in the SDW phase turned out to be useless for this purpose.
These scaling relations are only valid at temperatures $T\ll \Delta_c$ which
are not accessible by TMRG because the charge gap $\Delta_c$ is already very
small near the KT transition. We could, however, show that extracting the spin
gap from the magnetic susceptibility where it is large enough and fitting it
to a field theory formula does allow to determine the transition line
reasonably well. In particular, the results clearly confirm that the two
transition lines do not coincide and that an intermediate phase exists. By
studying correlation lengths and static susceptibilities we confirmed that
this additional phase has long-range bond order. We were also able to
determine the tricritical point accurately and found $U_t=6.7\pm 0.2$,
$V_t=3.5\pm 0.1$. Furthermore, we showed that above this point long-range bond
order does not exist. Instead we find that BOW correlations can be dominant in
this regime while still decaying algebraically at zero temperature. The
resulting phase diagram is in good quantitative agreement with the most recent
zero temperature DMRG study.\cite{Zhang} However, it does not agree with the
phase diagram found in Ref.~\onlinecite{SandvikBalents} with the BOW phase
existing even above the tricritical point.

\acknowledgments The authors acknowledge helpful discussions with
E.~Jeckelmann and S.~Nishimoto. This work has been supported by the DFG
Schwerpunkt SP1073 and Graduiertenkolleg GK1052 (S.G., A.K.).


\begin{thebibliography}{21}
\expandafter\ifx\csname natexlab\endcsname\relax\def\natexlab#1{#1}\fi
\expandafter\ifx\csname bibnamefont\endcsname\relax
  \def\bibnamefont#1{#1}\fi
\expandafter\ifx\csname bibfnamefont\endcsname\relax
  \def\bibfnamefont#1{#1}\fi
\expandafter\ifx\csname citenamefont\endcsname\relax
  \def\citenamefont#1{#1}\fi
\expandafter\ifx\csname url\endcsname\relax
  \def\url#1{\texttt{#1}}\fi
\expandafter\ifx\csname urlprefix\endcsname\relax\def\urlprefix{URL }\fi
\providecommand{\bibinfo}[2]{#2}
\providecommand{\eprint}[2][]{\url{#2}}

\bibitem[{\citenamefont{Hirsch}(1984)}]{Hirsch_EHM}
\bibinfo{author}{\bibfnamefont{J.~E.} \bibnamefont{Hirsch}},
  \bibinfo{journal}{Phys. Rev. Lett.} \textbf{\bibinfo{volume}{53}},
  \bibinfo{pages}{2327} (\bibinfo{year}{1984}).

\bibitem[{\citenamefont{Bari}(1971)}]{Bari}
\bibinfo{author}{\bibfnamefont{R.~A.} \bibnamefont{Bari}},
  \bibinfo{journal}{Phys. Rev. B} \textbf{\bibinfo{volume}{3}},
  \bibinfo{pages}{2662} (\bibinfo{year}{1971}).

\bibitem[{\citenamefont{van Dongen}(1994)}]{vanDongen}
\bibinfo{author}{\bibfnamefont{P.~G.~J.} \bibnamefont{van Dongen}},
  \bibinfo{journal}{Phys. Rev. B} \textbf{\bibinfo{volume}{49}},
  \bibinfo{pages}{7904} (\bibinfo{year}{1994}).

\bibitem[{\citenamefont{Nakamura}(2000)}]{NakamuraPRB}
\bibinfo{author}{\bibfnamefont{M.}~\bibnamefont{Nakamura}},
  \bibinfo{journal}{Phys. Rev. B} \textbf{\bibinfo{volume}{61}},
  \bibinfo{pages}{16377} (\bibinfo{year}{2000}).

\bibitem[{\citenamefont{S\'olyom}(1979)}]{Solyom}
\bibinfo{author}{\bibfnamefont{J.}~\bibnamefont{S\'olyom}},
  \bibinfo{journal}{Adv. Phys.} \textbf{\bibinfo{volume}{28}},
  \bibinfo{pages}{201} (\bibinfo{year}{1979}).

\bibitem[{\citenamefont{Nakamura}(1999)}]{NakamuraJPSJ}
\bibinfo{author}{\bibfnamefont{M.}~\bibnamefont{Nakamura}},
  \bibinfo{journal}{J. Phys. Soc. Jpn.} \textbf{\bibinfo{volume}{68}},
  \bibinfo{pages}{3123} (\bibinfo{year}{1999}).

\bibitem[{\citenamefont{Sengupta et~al.}(2002)\citenamefont{Sengupta, Sandvik,
  and Campell}}]{SenguptaSandvik}
\bibinfo{author}{\bibfnamefont{P.}~\bibnamefont{Sengupta}},
  \bibinfo{author}{\bibfnamefont{A.~W.} \bibnamefont{Sandvik}},
  \bibnamefont{and} \bibinfo{author}{\bibfnamefont{D.~K.}
  \bibnamefont{Campell}}, \bibinfo{journal}{Phys, Rev. B}
  \textbf{\bibinfo{volume}{65}}, \bibinfo{pages}{155113}
  (\bibinfo{year}{2002}).

\bibitem[{\citenamefont{Sandvik et~al.}(2004)\citenamefont{Sandvik, Balents,
  and Campbell}}]{SandvikBalents}
\bibinfo{author}{\bibfnamefont{A.~W.} \bibnamefont{Sandvik}},
  \bibinfo{author}{\bibfnamefont{L.}~\bibnamefont{Balents}}, \bibnamefont{and}
  \bibinfo{author}{\bibfnamefont{D.~K.} \bibnamefont{Campbell}},
  \bibinfo{journal}{Phys. Rev. Lett.} \textbf{\bibinfo{volume}{92}},
  \bibinfo{pages}{236401} (\bibinfo{year}{2004}).

\bibitem[{\citenamefont{Tsuchiizu and Furusaki}(2002)}]{TsuchiizuFurusaki}
\bibinfo{author}{\bibfnamefont{M.}~\bibnamefont{Tsuchiizu}} \bibnamefont{and}
  \bibinfo{author}{\bibfnamefont{A.}~\bibnamefont{Furusaki}},
  \bibinfo{journal}{Phys. Rev. Lett.} \textbf{\bibinfo{volume}{88}},
  \bibinfo{pages}{056402} (\bibinfo{year}{2002}).

\bibitem[{\citenamefont{Jeckelmann}(2002)}]{Jeckelmann}
\bibinfo{author}{\bibfnamefont{E.}~\bibnamefont{Jeckelmann}},
  \bibinfo{journal}{Phys. Rev. Lett.} \textbf{\bibinfo{volume}{89}},
  \bibinfo{pages}{236401} (\bibinfo{year}{2002}).

\bibitem[{\citenamefont{Zhang}(2004)}]{Zhang}
\bibinfo{author}{\bibfnamefont{Y.~Z.} \bibnamefont{Zhang}},
  \bibinfo{journal}{Phys. Rev. Lett.} \textbf{\bibinfo{volume}{92}},
  \bibinfo{pages}{246404} (\bibinfo{year}{2004}).

\bibitem[{\citenamefont{Tam et~al.}(2006)\citenamefont{Tam, Tsai, and
  Campbell}}]{TamTsai}
\bibinfo{author}{\bibfnamefont{K.-M.} \bibnamefont{Tam}},
  \bibinfo{author}{\bibfnamefont{S.-W.} \bibnamefont{Tsai}}, \bibnamefont{and}
  \bibinfo{author}{\bibfnamefont{D.~K.} \bibnamefont{Campbell}},
  \bibinfo{journal}{Phys. Rev. Lett.} \textbf{\bibinfo{volume}{96}},
  \bibinfo{pages}{036408} (\bibinfo{year}{2006}).

\bibitem[{\citenamefont{I.Peschel et~al.}(1999)\citenamefont{I.Peschel, Wang,
  Kaulke, and Hallberg}}]{Peschel}
\bibinfo{editor}{\bibnamefont{I.Peschel}},
  \bibinfo{editor}{\bibfnamefont{X.}~\bibnamefont{Wang}},
  \bibinfo{editor}{\bibfnamefont{M.}~\bibnamefont{Kaulke}}, \bibnamefont{and}
  \bibinfo{editor}{\bibfnamefont{K.}~\bibnamefont{Hallberg}}, eds.,
  \emph{\bibinfo{title}{Density-Matrix Renormalization, Lecture Notes in
  Physics}}, vol. \bibinfo{volume}{528} (\bibinfo{publisher}{Springer},
  \bibinfo{address}{Berlin}, \bibinfo{year}{1999}), \bibinfo{note}{and
  references therein}.

\bibitem[{\citenamefont{Glocke et~al.}(2006)\citenamefont{Glocke, Kl\"umper,
  and Sirker}}]{GlockeKluemperSirker}
\bibinfo{author}{\bibfnamefont{S.}~\bibnamefont{Glocke}},
  \bibinfo{author}{\bibfnamefont{A.}~\bibnamefont{Kl\"umper}},
  \bibnamefont{and} \bibinfo{author}{\bibfnamefont{J.}~\bibnamefont{Sirker}},
  \bibinfo{journal}{cond-mat/0610689}  (\bibinfo{year}{2006}).

\bibitem[{\citenamefont{Sirker and
  Kl\"umper}(2002{\natexlab{a}})}]{SirkerKluemperEPL}
\bibinfo{author}{\bibfnamefont{J.}~\bibnamefont{Sirker}} \bibnamefont{and}
  \bibinfo{author}{\bibfnamefont{A.}~\bibnamefont{Kl\"umper}},
  \bibinfo{journal}{Europhys. Lett.} \textbf{\bibinfo{volume}{60}},
  \bibinfo{pages}{262} (\bibinfo{year}{2002}{\natexlab{a}}).

\bibitem[{\citenamefont{J\"uttner et~al.}(1998)\citenamefont{J\"uttner,
  Kl\"umper, and Suzuki}}]{JuettnerKluemper_Hubbard}
\bibinfo{author}{\bibfnamefont{G.}~\bibnamefont{J\"uttner}},
  \bibinfo{author}{\bibfnamefont{A.}~\bibnamefont{Kl\"umper}},
  \bibnamefont{and} \bibinfo{author}{\bibfnamefont{J.}~\bibnamefont{Suzuki}},
  \bibinfo{journal}{Nucl. Phys. B} \textbf{\bibinfo{volume}{522}},
  \bibinfo{pages}{471} (\bibinfo{year}{1998}).

\bibitem[{\citenamefont{J\"uttner et~al.}(1997)\citenamefont{J\"uttner,
  Kl\"umper, and Suzuki}}]{JuettnerKluemper}
\bibinfo{author}{\bibfnamefont{G.}~\bibnamefont{J\"uttner}},
  \bibinfo{author}{\bibfnamefont{A.}~\bibnamefont{Kl\"umper}},
  \bibnamefont{and} \bibinfo{author}{\bibfnamefont{J.}~\bibnamefont{Suzuki}},
  \bibinfo{journal}{Nucl. Phys. B} \textbf{\bibinfo{volume}{487}},
  \bibinfo{pages}{650} (\bibinfo{year}{1997}).

\bibitem[{\citenamefont{Nakamura et~al.}(1999)\citenamefont{Nakamura, Kitazawa,
  and Nomura}}]{NakamuraKitazawa}
\bibinfo{author}{\bibfnamefont{M.}~\bibnamefont{Nakamura}},
  \bibinfo{author}{\bibfnamefont{A.}~\bibnamefont{Kitazawa}}, \bibnamefont{and}
  \bibinfo{author}{\bibfnamefont{K.}~\bibnamefont{Nomura}},
  \bibinfo{journal}{Phys. Rev. B} \textbf{\bibinfo{volume}{60}},
  \bibinfo{pages}{7850} (\bibinfo{year}{1999}).

\bibitem[{\citenamefont{Vollhardt}(1997)}]{Vollhardt}
\bibinfo{author}{\bibfnamefont{D.}~\bibnamefont{Vollhardt}},
  \bibinfo{journal}{Phys. Rev. Lett.} \textbf{\bibinfo{volume}{78}},
  \bibinfo{pages}{1307} (\bibinfo{year}{1997}).

\bibitem[{\citenamefont{Kemper and
  Schadschneider}(2003)}]{KemperSchadschneider}
\bibinfo{author}{\bibfnamefont{A.}~\bibnamefont{Kemper}} \bibnamefont{and}
  \bibinfo{author}{\bibfnamefont{A.}~\bibnamefont{Schadschneider}},
  \bibinfo{journal}{Phys. Rev. B} \textbf{\bibinfo{volume}{68}},
  \bibinfo{pages}{235102} (\bibinfo{year}{2003}).

\bibitem[{\citenamefont{Sirker and
  Kl\"umper}(2002{\natexlab{b}})}]{SirkerKluemperPRB}
\bibinfo{author}{\bibfnamefont{J.}~\bibnamefont{Sirker}} \bibnamefont{and}
  \bibinfo{author}{\bibfnamefont{A.}~\bibnamefont{Kl\"umper}},
  \bibinfo{journal}{Phys. Rev. B} \textbf{\bibinfo{volume}{66}},
  \bibinfo{pages}{245102} (\bibinfo{year}{2002}{\natexlab{b}}).

\end{thebibliography}
\end{document}